\documentclass[twoside,11pt]{article}

\usepackage[dvips]{graphicx}
\usepackage{setspace}

\begin{document}

\pagenumbering{roman}
\newtheorem{theorem}{\sc\bf Theorem}
\newtheorem{lemma}[theorem]{\sc\bf Lemma}
\newtheorem{corollary}[theorem]{\sc\bf Corollary}
\newtheorem{proposition}[theorem]{\sc\bf Proposition}
\newcommand{\boldgreek}[1]{\mbox{\boldmath $#1$}}

\title{\vspace{-2.0cm}
\Large On Pinsker's Type Inequalities and Csisz\'{a}r's $f$-divergences. \\
Part I: Second and Fourth-Order Inequalities.}
\author{{\normalsize\sc Gustavo L.\ Gilardoni}
\thanks{{\em Address for correspondence}:
Departamento de Estat{\'\i}stica, Universidade de Bras{\'\i}lia,
Bras{\'\i}lia, DF 70910--900, Brazil. E-mail: gilardon@unb.br.
Tel/Fax: +5561 3273-6317}
\\
{\normalsize\em Universidade de Bras{\'\i}lia}
\date{}
}
\maketitle

\vspace{-1.0cm}

\begin{abstract}
We study conditions on $f$ under which
an $f$-divergence $D_f$ will satisfy
$D_f \geq c_f V^2$
or $D_f \geq c_{2,f} V^2 + c_{4,f} V^4$, where
$V$ denotes variational distance and
the coefficients $c_f$, $c_{2,f}$ and $c_{4,f}$ are {\em best possible}.
As a consequence, we obtain lower bounds in terms of $V$ for many
well known distance and divergence measures.
For instance, let
$D_{(\alpha)} (P,Q) =
[\alpha (\alpha-1)]^{-1} [\int q^{\alpha} p^{1-\alpha} \, d \mu -1]$
and
${\cal I}_\alpha (P,Q) =
(\alpha -1)^{-1} \log [\int p^\alpha q^{1-\alpha} \, d \mu]$
be respectively the {\em relative information of type} ($1-\alpha$)
and {\em R\'{e}nyi's information gain of order} $\alpha$.
We show that
$D_{(\alpha)} \geq \frac{1}{2} V^2 +
\frac{1}{72} (\alpha+1)(2-\alpha) V^4$ whenever
$-1 \leq \alpha \leq 2$, $\alpha \not= 0,1$
and that
${\cal I}_{\alpha} \geq \frac{\alpha}{2} V^2 + \frac{1}{36} \alpha (1 + 5 \alpha - 5 \alpha^2 ) V^4$
for  $0 < \alpha < 1$.
Pinsker's inequality $D \geq \frac{1}{2} \, V^2$ and its
extension $D \geq \frac{1}{2} \, V^2 + \frac{1}{36} \, V^4$ are
special cases of each one of these.

\vspace{0.5cm}

\vspace{0.2cm}


\vspace{1.5cm} 
\hfill \today

\end{abstract}

\newpage
\pagenumbering{arabic}

\section{Introduction}

Studying the relationship among the information
divergence $D$
%$D(P,Q) = \int p \log (p/q) \, d \mu$
and the variational distance $V$
%$V (P,Q) = \int |q-p| \, d \mu$
%(here and throughout,
%$P$ and $Q$ are probability measures on a measurable space
%$(\Omega , {\cal A})$ and $p$ and $q$ are their
%densities or Radon-Nikodym derivatives
%with respect to a common dominating measure $\mu$)
or, more specifically,
determining lower bounds on $D$ in terms of $V$,
has been of interest at least since 1959, when
Volkonskij and Rozanov \cite{VoRo59}
showed that $D \geq V - \log ( 1+V)$.
The best known result in this direction is usually referred to
as Pinsker's inequality and states that $D \geq \frac{1}{2} V^2$.
In general, studying the relationship between $D$ and $V$
is important because it allows one to
"... translate results from information theory (results involving $D$)
to results in probability theory (results involving $V$) and {\em vice versa}"
(Fedotov, Harremo{\"e}s and Tops{\o}e, \cite{fedotovetal03}).
For instance, Barron \cite{barron86} found a strengthened version
of the central limit theorem by showing convergence in the
sense of relative entropy and then using Pinsker's inequality
to conclude convergence in the variational norm.
In different settings, this idea
has also been used by
Tops{\o}e \cite{topsoe79} and Harremo{\"e}s and Ruzankin \cite{harruzankin04}.
Interestingly, these kind of results and its relation with
Gagliardo-Nirenberg and generalized Sobolev inequalities
have been used recently in order to obtain the decay rate
of solutions of nonlinear diffusion equations---see
Del Pino and Dolbeault \cite{delpino02} and references therein.
%Carrillo, J{\"u}ngel, Markowich, Toscani and Unterreiter \cite{carrilloetal01} and
%Unterreiter, Arnold, Markowich and Toscani \cite{unteretal00}
%among others.

Pinsker's inequality was proved independently by
Csisz\'{a}r \cite{Csiszar67} and Kemperman \cite{Kemperman69},
building on previous work by
Pinsker \cite{Pinsker64}, Kean \cite{Kean66} and Csisz\'{a}r \cite{Csiszar66}.
The constant $\frac{1}{2}$ in $D \geq \frac{1}{2} \, V^2$ is {\em best
possible}, in the sense that there is
a probability space and two sequences of probability measures $P_n$ and $Q_n$
such that $D(P_n, Q_n)/ V^2(P_n,Q_n) \downarrow \frac{1}{2}$.
Sharpened Pinsker type
inequalities bounding $D$ by higher-order polynomials in $V^2$ are
also available. For instance, $D \geq \frac{1}{2} V^2 +
\frac{1}{36} V^4$---
see Kullback \cite{Kullback67,Kullback70} and Vajda \cite{Vajda70}, where again the constant
$\frac{1}{36}$ is best possible, in the sense that there are
sequences $P_n$ and $Q_n$ such that
$[D(P_n,Q_n ) -\frac{1}{2} V^2 (P_n , Q_n)] /
V^4 (P_n , Q_n) \downarrow \frac{1}{36}$.
More recently, Tops{\o}e showed
in \cite{topsoe2e100} that $D \geq \frac{1}{2} V^2 +\frac{1}{36}
V^4 + \frac{1}{270} V^6 + \frac{221}{340200} V^8$,
while Fedotov {\em et al}
\cite{fedotovetal03} have obtained a parametrization of the curve
$v \mapsto L(v) = \inf \{ D(P,Q) \!: \, V(P,Q) = v \}$ in terms
of hyperbolic trigonometric functions and argue in
\cite{fedotovetal03b} that the best possible extended Pinsker
inequality contains terms up to and including $V^{48}$.

Let $P$ and $Q$ be probability measures on a measurable space
$(\Omega , {\cal A})$ and $p$ and $q$ their
densities or Radon-Nikodym derivatives
with respect to a common dominating measure $\mu$.
The information divergence is
$D(P,Q) = \int p \log (p/q) \, d \mu$ and
is also known as
{\em relative entropy} or {\em Kullback-Leibler divergence}.
The variational (or $L^1$) distance is
$V (P,Q) = \int |q-p| \, d \mu$.
The $f$-divergence generated
by $f$ is $D_f (P,Q) = \int p f (q/p) \, d \mu$,
where $f: (0,\infty) \rightarrow \mbox{\boldmath$R$}$
is convex and $f(1) =0$.
Jensen's inequality implies that $D_f (P,Q) \geq 0$
with equality holding if and only if $P=Q$, provided
that $f$ is strictly convex at $u=1$.
Hence, $D_f (P,Q)$ can be thought of as a
measure of discrepancy between $P$ and $Q$.
The class of $f$-divergences was introduced by
Csisz{\'a}r \cite{Csiszar67} and Ali and Silvey \cite{Ali66}.
It includes many of the most popular distances and discrepancy
measures between probability measures.
Both $D$ and $V$ belong to this class, respectively
for $f(u) = - \log u$ and $f(u) = |u-1|$.
All of the following are also $f$-divergences:
the $\chi^2$ divergence
$\chi^2 (P, Q) = \int {(q - p )^2 \over p} \, d \mu =
\int {\,\, q^2 \over p} \, d \mu  -1$,
the Hellinger discrimination
$h^2 (P,Q) = \frac{1}{2} \int (\sqrt{q} -\sqrt{p})^2 \, d \mu = 1 - \int \sqrt{qp} \, d \mu$,
the Triangular or Harmonic divergence
$\Delta (P,Q) = \int {(q - p )^2 \over \! p + q} \, d \mu =
4 \int {q^2 \over p + q}  \, d \mu  -2 =
2 - 4 \int {p q \over p + q}  \, d \mu $,
the Capacitory discrimination
$C(P,Q) = D(P, M) + D(Q,M)$ where $M = (P+Q)/2$  and
the Jeffrey's divergence
$J(P,Q) = D(P,Q) + D(Q,P) = \int (q-p) \log (q/p) \, d \mu$.
A convenient one-parameter family which includes many of the above
as special cases is
generated by the convex functions
$f(u) = [\alpha (\alpha-1)]^{-1} (u^{\alpha} -1)$ ($\alpha \not= 0,1$).
The resulting divergence
$D_{(\alpha)} (P,Q) = [\alpha (\alpha-1)]^{-1} [\int q^{\alpha} p^{1-\alpha} \, d \mu -1]$
is called {\em relative information of type} ($1 - \alpha$)
by Vajda \cite{vajda89} and Taneja \cite{taneja04}.
It is easy to check that $\chi^2 = 2 D_{(2)}$,
$4 h^2 = D_{(1/2)}$ and $D = \lim_{\alpha \rightarrow 0} D_{(\alpha)}$.
The Tsallis' and the Cressie-Read divergences,
which are used extensively in many areas
including physics, economics and statistics,
are respectively 
$T_{\alpha} = \alpha D_{(1-\alpha)}$ and 
$CR_{\lambda} = D_{(- \lambda)}$
(see \cite{Golan02,Withers99}).
Finally, the R\'{e}nyi's {\em information gain of order $\alpha > 0$},
${\cal I}_\alpha (P,Q) = (\alpha -1)^{-1} \log [\int p^\alpha q^{1-\alpha} \, d \mu]$, of which
the information divergence is also a special case as $\alpha \rightarrow 1$,
although not itself an $f$-divergence,
can be expressed as
${\cal I}_{\alpha} = (\alpha -1)^{-1} \log [1-\alpha (1-\alpha) D_{(1-\alpha)}]$,
cf.\ \cite{Csiszar67,Golan02,Csis95}.
%the Tsallis divergence of order $\alpha$ ($f(u)=(1-\alpha)^{-1} (u^{1-\alpha}-1)$),
%$I^T_\alpha (P,Q) = (1-\alpha)^{-1} [\int p^{\alpha} q^{1-\alpha} -1]$ and
%
%the Cressie-Read divergence of order $\alpha$
%($f(u) = [\alpha (1+\alpha)]^{-1} (u^\alpha -1)$).

Regarding the relationship
between $f$-divergences and $V$,
bounds are available for some special cases
involving divergences which are
more or less easy to manipulate.
For instance, it is known that
$\chi^2 \geq V^2$,
$\frac{1}{2} V^2 \leq \Delta \leq V$ and
$\frac{1}{8} V^2 \leq h^2 \leq \frac{1}{2} V$,
see Dacunha-Castelle \cite{Dacunha-Castelle78},
LeCam \cite{Cam86},
Dragomir, Glu\u{s}\u{c}evi\'{c} and Pearce \cite{Dragomir01}
and Tops{\o}e \cite{topsoe2e100}.
A precise bound is available for the
Capacitory divergence, for which
Tops{\o}e \cite{Topsoe00ine}
showed that
$C(P,Q) \geq \sum_{n=1}^{\infty} [n (2n-1) 2^{2n}]^{-1} V^{2n}$
$= \log \frac{4-V^2}{4} + \frac{V}{2} \, \log \frac{2+V}{2-V}$.
%$= \frac{2-v}{2} \log \frac{2-v}{2} + \frac{2+v}{2} \log \frac{2+v}{2}$.

Although we know of no general result 
giving a lower bound for 
$f$-divergences in terms of $V$, 
it appears as intuitively clear to us 
from the fact that $f$-divergences share 
many of the properties of $D$ that 
inequalities similar to Pinsker's should also hold for
other divergences.
This should be the case, for instance,
of relative information of type ($1 - \alpha$) with
$\alpha$ close to zero or that of 
R\'{e}nyi's information gain of order $\alpha$ with
$\alpha$ close to one.
Maybe the closest to a general statement giving a kind
of lower bound for an arbitrary $D_f$ in terms of $V$
is in Csisz\'{a}r \cite[Theorem 1]{Csiszar67topological}, which
states that $D_f (P,Q) < \epsilon$ implies,
for sufficiently small $\epsilon$, that
$\frac{f''(1)}{2 \sqrt{2}} V^2 (P,Q) < \epsilon$
({\em cf.} our Theorem \ref{prop.pinskerf} below),
%(compare this to our Theorem \ref{prop.pinskerf} below),
implying then that $V$ should be small whenever $D_f$ is small enough.

This paper will be the first of a series dealing with
the relationship between $f$-divergences and variational distance.
In particular, our objective here is to discuss
conditions under which an $f$-divergence satisfies
either a Pinsker's type inequality $D_f \geq c_f V^2$
or a fourth-order inequality $D_f \geq c_{2,f} V^2 + c_{4,f} V^4$.
We will show in Section 3 that a sufficient condition
for $D_f \geq c_f V^2$ is that the ratio between $(u-1)^2$ and
the difference between the generating $f$ and its tangent at
$u=1$ be upper bounded by a straight line $a+bu$ with nonnegative
coefficients $a$ and $b$ such that $a+b = c_f^{-1} = \frac{2}{f''(1)}$,
if we want $c_f$ to be {\em best possible}.
A sufficient condition for having a fourth-order inequality,
always with {\em best possible} coefficients,
is presented in Section 4.
Each of these theorems is followed by a corollary which
gives conditions on the derivatives of $f$ which are
easier to check in practice than the original
conditions on $f$.
As a consequence of these we show in Section 3 that the
relative information of type ($1 - \alpha$) satisfies
$D_{(\alpha)} \geq \frac{1}{2} V^2$ whenever
$-1 \leq \alpha \leq 2$, $\alpha \not= 0,1$.
This inequality is improved in Section 4 to
$D_{(\alpha)} \geq \frac{1}{2} V^2 + \frac{1}{72} (\alpha+1) (2-\alpha) V^4$.
Using that
${\cal I}_{\alpha} = (\alpha -1)^{-1} \log [1-\alpha (1-\alpha) D_{(1-\alpha)}]$,
we also obtain that the R\'{e}nyi's information gain of order $\alpha$ satisfies
${\cal I}_{\alpha} \geq \frac{1}{2} \alpha \,V^2 + \frac{1}{36} \alpha (1 + 5 \alpha - 5 \alpha^2 ) \, V^4$
whenever $0 < \alpha < 1$.

Besides sections 3 and 4 dealing respectively with second and fourth-order
inequalities, the rest of the paper is organized as follows.
Section 2 introduces some additional notation and states
a fundamental inequality between powers of $V$ and $D_f$
in Corollary \ref{prop.lemma.vmn}.
In Section 5 we bring forward an argument from the sequel
of this paper and briefly discuss why the tools that we use here to
obtain second and fourth-order inequalities are insufficient
to obtain sixth and higher-order inequalities when we are
interested in {\em best possible} coefficients.
%Section 6 discusses a few questions which,
%although interesting to us, are left open by our approach.
Some technical results needed in Section 4 are
presented in an Appendix.
Finally, since some of the proofs in Section 4 and in the 
Appendix require somewhat lengthy calculations, 
we have recorded a MAPLE script that could help 
the reader interested in checking them.
Although the script is not included here for reason of space, 
it is available from us on request. 
Notwithstanding, we stress that we have included in the paper 
what we believe are full and complete proofs for all statements made.     

%In a sense, our tools are quite simple and are obtained by looking
%at proofs of the corresponding inequalities for the
%information divergence. For instance, we will show
%that Pinsker's inequality is a consequence of the inequality
%\begin{equation}
%\label{ineq.log.2}
%\log u \leq u - 1 - \frac{1}{2} \frac{(u-1)^2}{1+\frac{2}{3} (u-1)}
%\,\,\,\,\,\,\,\,\,\,\,\,\,\, u > 0 \,.
%\end{equation}
%(Incidentally, (\ref{ineq.log.2}) can be rewritten as
%$\log (1+x) \leq x (6+x)/2(3+2x)$, $x > -1$, which
%appears already in Tops{\o}e \cite{citarolog}, although
%holding only for $x>0$).

\section{Notation and preliminary considerations}

Throughout, equalities or inequalities between divergences will
be understood to hold for every pair of probability measures,
so that we will write for instance
"$D \geq {\displaystyle \frac{1}{2}} \, V^2$"
instead of
"$D(P,Q) \geq {\displaystyle \frac{1}{2}} \, V^2 (P,Q)$ for every $P,Q$".
An equivalent definition for the Variational Distance is
$V(P,Q) = 2 \, \sup \{ |Q(A) - P(A)| : \, A \in {\cal A} \} =
2 \, [Q(B) - P(B)]$, where
$B = \{ \omega \in \Omega : \, q(\omega) \geq p(\omega) \}$.
Hence $0 \leq V(P,Q) \leq 2$ with equality holding respectively
if and only if $P=Q$ or $P \perp Q$.
It is well known that
the Information Divergence satisfies
$0 \leq D(P,Q) \leq + \infty$.
$D(P,Q) =0$ can occur if and only if $P=Q$, while
$P \not\ll Q$ implies that $D(P,Q) = + \infty$, although the
reciprocal does not hold.

To avoid unnecessary discussion, we will assume the usual
conventions $f(0) = \lim_{u \downarrow 0} f(u)$,
$0 \cdot f(0/0) = 0$ and
$0 \cdot f(a/0) = \lim_{\epsilon \downarrow 0} \epsilon f (a / \epsilon) =
\lim_{u \uparrow + \infty} f(u)/u$.
An $f$-divergence does not determine univocally the
associated $f$.
Indeed, for any $a$ fixed, $D_f$ and $D_{f - a (u-1)}$ are identical.
For instance, $D = D_{- \log u} = D_{u-1 - \log u}$.
For any convex $f$ we will let $\tilde{f} (u) = f(u) - f'(1) (u-1)$,
which is nonnegative due to convexity considerations (more precisely,
$f'(1)$ can be taken to be any number between the left and right
derivatives of $f$ at $u=1$). We note here that second and 
higher-order derivatives of $f$ and $\tilde{f}$ coincide.
Indeed,  we will often switch from one to the other
in sections 3 and 4 .

In general, $f$-divergences are not symmetric,
in the sense that $D_f (P,Q)$ does not necessarily
equals $D_f (Q,P)$, unless the generating
$f$ satisfies that $f(u) = u f(1/u) + a (u-1)$ for
some fixed $a$. This is the case for instance
of $V$ and the $h^2$ and $\Delta$ divergences but
not that of $D$ or $\chi^2$. Whenever an $f$-divergence
is not symmetric we could define the {\em reversed} divergence
by letting $f_R (u) = u f(1/u)$, so that
$D_{f_R} (P,Q) = D_f (Q,P)$.
Similarly, beginning from an arbitrary $f$-divergence
it is possible to construct a symmetric measure by
using the convex function
$f_S (u) = f(u) + f_R (u)$.
For instance, the reversed information divergence is
$D_R (P,Q) = \int q \log (q/p) \, d \mu$ and
its symmetrized version is the already
mentioned Jeffrey's divergence.

The following lemma is slightly more general than what we will
actually need in Sections 3 and 4. It gives an upper bound on
$| \int g q \, d \mu - \int g p \, d \mu \, | = | E_Q g - E_P g \, |$
in terms of a certain higher moment of $g$ and an $f$-divergence $D_f$.
If we interpret $Q$ as an approximation to $P$, then
$| E_Q g - E_P g |$ is the error between the corresponding
approximate and actual expectations.
The lemma generalizes results which we have used in
\cite{gilardoni05,gilardoni06}
in order to obtain upper limits
for the approximating error in
the context of Bayesian Statistics.

\begin{lemma}
\label{prop.lemmagmn}
Let $g$ be both $P$ and $Q$ integrable, $k(u) \geq 0$ such that
$\int p \, k(q/p) \, d \mu =1$, and $n > 1$.
Then for any fixed $a$
\begin{equation}
\label{eqn.lemma.gmn}
\left| \int g  q \, d \mu - \int g  p \, d \mu  \, \right|^n \leq \;
\sup_{u>0, \, u \not= 1} \left\{ \frac{(u-1)^n}{\tilde{f} (u) k^{n-1} (u)} \right\} \cdot
[ {\rm E}_r |g-a|^{n/(n-1)} ]^{n-1} \cdot D_f (P,Q) \,,
\end{equation}
where $r = p \, k (q/p)$,
%$a = {\rm E}_r g = \int g r \, d \mu$,
${\rm E}_r |g-a|^{n/(n-1)} = \int |g-a|^{n/(n-1)} r \, d \mu$ is the
$[n/(n-1)]-$th moment of $g$ around $a$ with respect to the probability
density $r$ and, as before,
$\tilde{f} (u) = f(u) - f'(1) (u-1)$ and $f'(1)$ is a number between the left and the
right derivative of $f$ at $u=1$.
\end{lemma}

{\sc Proof}:
Let $m = n/(n-1)$ (so that $n$ and $m$ are conjugate) 
and $C = \{ \omega:\, q(\omega) \not= p(\omega) \}$.
Then we have for any real $a$ that
\begin{eqnarray*}
\lefteqn{\left| \int gq \, d \mu - \int gp  \, d \mu \, \right| =
\left| \int_C (g -a)(q-p)  \, d \mu \, \right|
\leq \int_C |g -a| \, |q-p| \, d \mu}  \\
& &  =
\int_C \left[ \frac{|g-a| \, |q/p-1|}{\tilde{f}^{1/n} (q/p)} \right]
\tilde{f}^{1/n}(q/p) \, p \, d \mu  \leq
\left[ \int_C \frac{|g-a|^m \, |q/p-1|^m }{\tilde{f}^{m/n}(q/p)}
\, p  \, d \mu \right]^{1/m}
\cdot \left[ D_f (P,Q) \right]^{1/n} \\
& & =
\left[ \int_C \frac{|g-a|^m \, |q/p-1|^m}{\tilde{f}^{m/n} (q/p) k(q/p)}
\, k(q/p) \, p  \, d \mu \right]^{1/m}
\cdot \left[ D_f (P,Q) \right]^{1/n} \\
& & \leq
\sup_{u > 0, u \not= 1} \left\{ \frac{(u-1)^m}{\tilde{f}^{m/n} (u) k(u)} \right\}^{1/m}
\; \cdot \left[ \int |g-a|^m r  \, d \mu \right]^{1/m}
\cdot \left[ D_f (P,Q) \right]^{1/n} \,,
\end{eqnarray*}
where the second inequality
follows from H\"{o}lder's inequality.
The desired result now follows
after taking the $n$-th power in the leftmost and
rightmost terms and noting that
$\sup_{u > 0, u\not= 1} \{ (u-1)^m / \tilde{f}^{m/n} (u) k(u) \}^{n/m} =
\sup_{u > 0, u \not= 1} \{ (u-1)^n / \tilde{f} (u) k^{n/m} (u) \} =
\sup_{u > 0, u \not= 1} \{ (u-1)^n / \tilde{f} (u) k^{n-1} (u) \}$.
\hfill \framebox{}

Taking $g = I_{p \geq q}$ and $a = \frac{1}{2}$,
the left hand side of (\ref{eqn.lemma.gmn})
becomes $2^{-n} |V(P,Q)|^n$, while
${\rm E}_r |g-a|^{n/(n-1)} = \int |I_{p \geq q} -\frac{1}{2}|^{n/(n-1)} r \, d \mu \leq 2^{-n/(n-1)}$.
Hence, we have the following corollary.

\begin{corollary}
\label{prop.lemma.vmn}
Let $k$, $f$ and $\tilde{f}$ be as before. Then
\begin{equation}
\label{eqn.lemma.vmn}
V^n (P,Q) \leq
\sup_{u > 0, u \not= 1} \left\{ \frac{(u-1)^n}{\tilde{f} (u) k^{n-1} (u)} \right\} \;
\cdot D_f (P,Q) \,
\end{equation}
\end{corollary}

{\sc Remark 1}. These results are still valid for
nonconvex $f$ provided that $\tilde{f} (u) \geq 0$ and
we interpret $D_f (P,Q) = \int \tilde{f} (q/p) \, p \, d \mu$.

{\sc Remark 2}. Although (\ref{prop.lemmagmn}) holds for any $a$,
we usually would like to use a value for which
${\rm E}_r |g-a|^{n/(n-1)}$ is small. Taking $a = \rm{E}_r g = \int g \, r \, d \mu$ could
be a good idea.

{\sc Remark 3}. A more precise formulation
would use the $L^{\infty}$ norm (or essential supremum) of
$(u-1)^n / \tilde{f} (u) k^{n-1}(u)$
with respect to the measure $r \, d \mu$
instead of the supremum for $u>0 , u \not= 1$.

%{\sc Remark 4}. Incluir comentario sobre igualdade
%na equacao com $V^n$.

Of particular interest will be the cases $n=2$ and
$n=4$, in which case equation (\ref{eqn.lemma.vmn}) becomes
\begin{equation}
\label{eqn.lemma.2}
V^2 (P,Q) \leq
\sup_{u > 0, u \not= 1} \left\{ \frac{(u-1)^2}{\tilde{f} (u) k(u)} \right\} \;
\; D_f (P,Q)
\end{equation}
and
\begin{equation}
\label{eqn.lemma.4}
V^4 (P,Q) \leq
\sup_{u > 0, u \not= 1} \left\{ \frac{(u-1)^4}{\tilde{f} (u) k^3 (u)} \right\} \;
\; D_f (P,Q) \,.
\end{equation}

Some interesting inequalities follow directly
from (\ref{eqn.lemma.2}).
For instance,
taking (i) $k(u)=1$ and $\tilde{f} (u) = (u-1)^2$ we
obtain that $\chi^2 \geq V^2$,
(ii) $k(u)=(1+u)/2$ and
$\tilde{f} (u) = (u-1)^2/(1+u)$, so that $(u-1)^2/\tilde{f}(u) k(u) \equiv 2$,
it follows that $\Delta \geq V^2$ and
(iii) $k(u) = (\sqrt{u}+1)^2/[2(2-h^2(P,Q))]$ and
$\tilde{f}(u) = \frac{1}{2} (\sqrt{u}-1)^2$, so that
$(u-1)^2/\tilde{f}(u) k(u) \equiv 4 [2 - h^2 (P,Q)]$,
we obtain that $4 h^2 (2- h^2) \geq V^2$---
see Kraft \cite{Kraft55}, cited in Dragomir et al. \cite{Dragomir01}.
Although we are not specially interested here in $f$-divergences
for which $f''(1) =0$, Corollary \ref{prop.lemmagmn} also gives some bounds
for this case. For instance, for the
Triangular Divergence of order $\nu >1$,
$\Delta_{\nu} (P,Q) = \int \frac{(q-p)^{2 \nu}}{(q+p)^{2 \nu -1}} \, d \mu$
(see Tops{\o}e \cite{Topsoe00ine} or
Dragomir et al. \cite{Dragomir01}),
we obtain after taking $n=2 \nu$,
$\tilde{f} (u) = (u-1)^{2 \nu} / (1+u)^{2 \nu -1}$ and
$k(u) = 1+u$ in (\ref{eqn.lemma.vmn})
that $\Delta_{\nu} \geq V^{2 \nu}$.

\section{Second-order inequalities}

\begin{figure}
    \centering
    \includegraphics[width=3.0in,height=4.0in,angle=270]{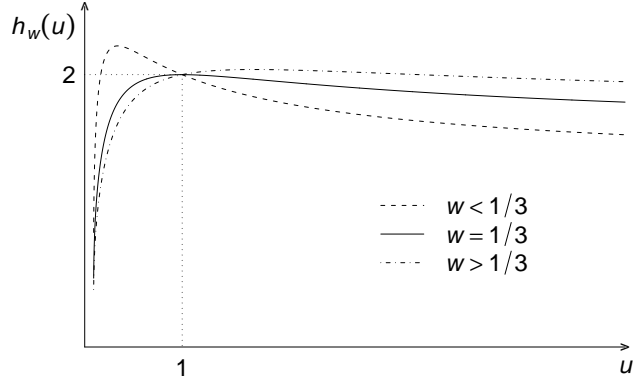}
    \caption{$h_w (u) = (u-1)^2 / (u-1- \log u) [w+(1-w) u]$ satisfies (i)
    $h_w (1) =2$ for every $w$ ( by continuity) and (ii) when $w$ is greater
    (smaller) than $\frac{1}{3}$, $h_w$ attains its maximum for a $u$ greater
    (respectively smaller) than $1$, hence for $w \not= \frac{1}{3}$,
    the maximum value of $h_w$ is greater than $2$.}
    \label{figura.hw}
\end{figure}

We first note that also Pinsker's inequality follows
from (\ref{eqn.lemma.2}).
To see this, take $\tilde{f} (u) = u-1-\log u$ and
$k(u) = (1 + 2u)/3$ to obtain that
$V^2 \leq \sup_{u>0, u \not= 1} h_{1/3} (u) \cdot D$, where
$h_{1/3} (u) = 3 (u-1)^2/[(u-1-\log u)(1+2u)]$
(the reason for the subindex $1/3$ here will be made
clear shortly).
Now note that $\sup_{u>0, u \not= 1} h_{1/3} (u) = 2$ because
$\lim_{u \to 1} h_{1/3} (u) =2$,while
$g(u) = 2 (u-1-\log u) (1+2u)-3 (u-1)^2 \geq 0$ for $u > 0$
since $g(1) = g'(1) =0$ and
$g''(u) = 2(u-1)^2 / u^2 \geq 0$.
Observe that using $k(u) = (1+2u)/3$ in the previous argument
amounts up to using the mixture
$k(q/p) \, p = \frac{1}{3} p + \frac{2}{3} q$ in the proof of
Lemma \ref{prop.lemmagmn}.
It is interesting to note the reason why
this and only this mixture works.
Using a mixture $wp + (1-w)q$
for some $0 \leq w \leq 1$, $w \not= \frac{1}{3}$
is equivalent to taking $k(u) = 1 + (1-w)(1-u)$
in equation (\ref{eqn.lemma.2}).
Now, define $h_w (u) = (u-1)^2 / (u-1- \log u) [1+(1-w) (u-1)]$
and observe that $\lim_{u \rightarrow 1} h_w (u) =2$
for any $w$, so that $\sup_{u>0, u \not= 1} h_w(u) \geq 2$,
while $h_w (u)$ attains its maximum value for $u=1$
if and only if $w=\frac{1}{3}$.
(see Figure \ref{figura.hw}).
Hence, it follows that $\sup_{u>0, u \not= 1} h_w(u) > 2$
whenever $w \not= \frac{1}{3}$ and 
using any
mixture with $w \not= \frac{1}{3}$ in Lemma \ref{prop.lemma.vmn}
will produce a less than optimal inequality
$D \geq c \, V^2$ with $c < \frac{1}{2}$.

The idea in the previous paragraph can
be formulated for arbitrary $f$-divergences.
Let $h_w (u) = (u-1)^2 / \{\tilde{f} (u) [(1+(1-w)(u-1)] \}$ and 
$b_w (u) = 1/h_w (u)$
and suppose that
$f(u) = f'(1) (u-1) + \frac{1}{2} f''(1) (u-1)^2 + \frac{1}{6} f'''(1) (u-1)^3 + o(|u-1|^3)$ with $f''(1) \not= 0$.
Equation (\ref{eqn.lemma.2}) implies now that
$D_f \geq [\sup_{u>0, u \not= 1} h_w (u)]^{-1} V^2$ for every $0 \leq w \leq 1$.
Note that $\lim_{u \to 1} h_w (u) =2/f''(1)$ for every $w$,
so that $[\sup_{u>0, u \not= 1} h_w (u)]^{-1} \geq f''(1)/2$.
In order to obtain the inequality $D_f \geq \frac{f''(1)}{2} V^2$,
we must find a $w=w_f$ such that $[\sup_{u>0, u \not= 1} h_w (u)]^{-1} = f''(1)/2$.
In other words, the (continuity corrected) function
$h_{w_f}$ should be maximized at $u=1$, or equivalently
$b_{w_f} = 1/h_{w_f}$ should be minimized at $u=1$.
It is easy to check that
$b_w (u) = \frac{f''(1)}{2} + [\frac{f''(1)}{2} (1-w) + \frac{f'''(1)}{6}] (u-1) +
o(|u-1|)$. Hence, for
$b_{w_f}$ to have a minimum at $u=1$, the
first-order term should vanish and hence
$w_f = 1 + \frac{1}{3} \frac{f'''(1)}{f''(1)}$.
Finally, for $h_{w_f} (u)$ to be actually maximized at $u=1$, we must have
that $h_{w_f} (u) \leq h_{w_f} (1) = 2/f''(1)$ and hence that
\begin{equation}
\label{cond.2nd}
\tilde{f} (u) [1 +(1-w_f) (u-1) ] \geq
\frac{f''(1)}{2} (u-1)^2 \,.
\end{equation}
This leads to the following theorem.

\begin{theorem}
\label{prop.pinskerf}
{\bf (Pinsker type inequality for $f$-divergences)}.
Suppose that the convex function $f$ is differentiable
up to order 3 at $u=1$ with $f''(1)>0$, and let
$w_f = 1 + \frac{1}{3} \frac{f'''(1)}{f''(1)}$. Then
(\ref{cond.2nd}) implies that $D_f \geq \frac{f''(1)}{2} V^2$.
The constant $\frac{f''(1)}{2}$ is best possible.
\end{theorem}

{\sc Proof:}
Although a rigorous proof can be obtained following the ideas above,
the following argument is easier once the right condition
has been identified.
First, note that for (\ref{cond.2nd}) to hold
we must have that $[1 +(1-w_f) (u-1) ] > 0$
for every $u \not= 1$.
% and hence that $-3 f''(1) \leq f'''(1) \leq 0$.
Hence, (\ref{cond.2nd}) implies that
\[
D_f (P,Q) = D_{\tilde{f}} (P,Q) \geq  \frac{f''(1)}{2}
\int \frac{(q/p-1)^2}{1+(1-w_f) (q/p-1)} \, p \, d \mu
\geq \frac{f''(1)}{2} \, V^2 (P,Q) \,,
\]
where the last inequality follows from (\ref{eqn.lemma.2}) after
taking $\tilde{f} (u) = (u-1)^2 / [1 +(1-w_f) (u-1) ]$
and $k(u) = [1 +(1-w_f) (u-1) ]$.

To show that $c_f$ is best possible, consider
a binary space and suppose that
$P$ assigns probabilities $p > 0$ and $(1-p) > 0$ to each
point of $\Omega$, say $P = (p,1-p)$.
For small $v$ define
$Q_v = (p+v/2, 1-p-v/2)$ so that
$V(P,Q_v) = v$ and
$D_f (P,Q_v) = p f(1+v/2p) + (1-p) f (1-v/2(1-p)) =
p \, \frac{f''(1)}{2} \, (v/2p)^2 +
(1-p) \, \frac{f''(1)}{2} \, [v/2(1-p)]^2 + o(v^2) =
\frac{f''(1)}{2} \, [4p(1-p)]^{-1} \, v^2 + o(v^2)$.
%p \, \frac{f''(1)}{2} \frac{v^2}{4 p^2} +
%(1-p) \frac{f''(1)}{2} \frac{v^2}{4 (1-p)^2} + o(v^2) =
%\frac{f''(1)}{2} \frac{1}{4 p(1-p)} \, v^2 + o(v^2)$.
Hence $\lim_{v \to 0} D_f (P,Q_v) /V^2(P,Q_v) =
\frac{f''(1)}{2} [4 p(1-p)]^{-1}$.
Taking $p = \frac{1}{2}$ completes the proof.
%To show that $\frac{f''(1)}{2}$ is best possible,
%consider a probability distribution $P$ and a set $A$
%such that $P(A) > 0$, and let
%$q_{\delta} = (1-\delta) p + [\delta / P(A)] p I_A$, where $I_A$
%is the indicator of $A$ and $\delta$ is sufficiently small so that
%$q_{\delta} \geq 0$.
%Then
%$V(P,Q_\delta) = \int \delta |I_A/P(A) -1| p \, d \mu =
%\int_A \delta \{[1-P(A)]/P(A) \} p \, d \mu +
%\int_{A^c} \delta p \, d \mu =
%2 \delta [1-P(A)]$ and
%$D_f(P, Q_\delta ) = D_{\tilde{f}} (P,Q_\delta) =
%\int_A f(1+ \delta [1-P(A)]/P(A) ) p \, d \mu +
%\int_{A^c} f(1- \delta) p \, d \mu =
%P(A) f(1+ \delta [1-P(A)]/P(A) ) + [1-P(A)] f(1- \delta)$.
%Hence for small
%$\delta$ we have that $D_f (P,Q_\delta) =
%P(A) \frac{f''(1)}{2} \{ \delta [1-P(A)]/P(A) \}^2 +
%[1-P(A)] \frac{f''(1)}{2} \delta^2 + o(\delta^2) =
%\frac{f''(1)}{2} \frac{1-P(A)}{P(A)} \,\delta^2 + o(\delta^2)$
%and
%$\lim_{\delta \to 0} \frac{D_f (P,Q_\delta)}{V^2(P,Q_\delta)} =
%\frac{f''(1)}{2} \frac{1}{4 P(A) [1-P(A)]}$.
%Taking $P(A) = \frac{1}{2}$ completes the proof.
\hfill \framebox{}

{\sc Remark.}
The last part of the proof suggests that,
when there is no set $A$ with $P(A) = \frac{1}{2}$,
a better constant
$c_f (P) = \frac{f''(1)}{2} [4 \,
\sup \{ P(A) [1-P(A)]: \, A \in {\cal A} \}]^{-1} > \frac{f''(1)}{2}$ can be
found so that $D_f (P,Q) \geq c_f (P) \, V^2 (P,Q)$ for every $Q$.
This problem has been addressed recently for the
information divergence by
Ordentlich and Weinberger \cite{Ordentlich05}).

%{\sc Remark 2.}
%In general $f$-divergences are not symmetric in their arguments,
%in the sense that $D_f(P,Q)$ and $D_f (Q,P)$ are not equal.
%In fact, $D_f (Q,P) = D_{f_R} (P,Q)$ where $f_R (u) = u f(1/u)$.
%Due to the symmetry of $V$, it is immediate that
%$D_f \geq \frac{f''(1)}{2} V^2$ implies then that also
%$D_{f_R} \geq \frac{f''(1)}{2} V^2$, but it remains
%the question whether applying theorem \ref{prop.pinskerf} we could find
%a better bound either for $D_{f_R}$ or even for
%the symmetrized divergence $D_f + D_{f_R}$ generated by
%the convex function $f(u) + u f(1/u)$.
%That this is not the case can be seen from the fact that
%$f''(1) = f^{''}_R (1)$ and that condition (\ref{cond.2nd})
%applied to $f$ and to $f_R$ gives equivalent inequalities.

For most divergences the condition in the next proposition
is easier to check than (\ref{cond.2nd}).

\begin{corollary}
\label{prop.pinskerf.easy}
Let $f$ and $\tilde{f}$ be as before,
$w_f = 1 + \frac{1}{3} \frac{f'''(1)}{f''(1)}$,
and suppose
that $f$ is three times differentiable with
$f''(u) > 0$ for all $u$. Then
\begin{equation}
\label{cond.2nd.easy}
{\rm sgn} (u-1) \left\{
\frac{f'''(u)}{f''(u)} [1 + (1-w_f)(u-1) ] + 3 (1 - w_f) \right\}
\geq 0
\end{equation}
implies (\ref{cond.2nd}) and hence that
$D_f \geq \frac{f''(1)}{2} \, V^2$.
\end{corollary}

{\sc Proof:}
Let $g(u) = \tilde{f} (u) [1+(1-w_f)(u-1)]-\frac{f''(1)}{2} (u-1)^2$.
Now $g(1) =0$,
$g'(u) = \tilde{f}' (u) [1+(1-w_f)(u-1)] + (1-w_f) \tilde{f} (u) - f''(1) (u-1)$
and hence $g'(1) =0$,
$g''(u)= \tilde{f}'' (u) [1+(1-w_f)(u-1)] + 2 (1-w_f) \tilde{f}' (u) -f''(1)$
and therefore $g''(1) = 0$ and finally
$g'''(u) = f''' (u) [1+(1-w_f)(u-1)] + 3 (1-w_f) f'' (1) (u)$.
%Since it follows from the definition of $f$ and $\tilde{f}$ that their
%second and higher-order derivatives coincide,
Hence, (\ref{cond.2nd.easy}) implies that
$g'''(u) \leq 0$ for $u<1$ and $g'''(1) \geq 0$ for $u>1$,
so the following Lemma implies
that $g$ must be nonnegative, which is equivalent to (\ref{cond.2nd}).
\hfill \framebox{}

{\sc Remark}.
Since we also have that $g'''(1) =0$,
(\ref{cond.2nd}) is also implied if
$g^{(4)} (u) = f^{(4)} [1 + (1-w_f)(u-1)] + 4 (1-w_f) f''' (u) \geq 0$,
but we usually find (\ref{cond.2nd.easy}) easier to check.

\begin{lemma}
\label{lemma.derivs}
Let $n \geq 1$ and $g:(0, \infty) \rightarrow \mbox{\boldmath $R$}$ be $(n+1)$ times
differentiable with $g(1) = g'(1) = \cdots = g^{(n)} (1) =0$, 
where $g^{(n)}$ is the n-th derivative of $g$,
and suppose that either (i) $n$ is even and
$g^{(n+1)} (u) \leq 0$ for $u < 1$ and $g^{(n+1)} (u) \geq 0$ for $u > 1$
or (ii) $n$ is odd and $g^{(n+1)} (u) \geq  0$ for every $u$.
Then $g(u) \geq 0$ for every $u$.
\end{lemma}

{\sc Proof}:
We prove first the case that $n$ is odd.
Since $g^{(n+1)} (u) \geq  0$ it follows that
$g^{(n-1)}$ is convex, and since $g^{(n-1)} (1) = (g^{(n-1)})' (1) =0$,
it must have a minimum at $u=1$, hence must be nonnegative. Repeat
the argument backwards to obtain that $g^{(n-3)}, \ldots , g$
are also nonnegative.
Now in the case (i) that $n$ is even, note that
$g^{(n)}$ decreases for $u<1$ and increases for $u>1$ and,
since $g^{(n)} (1) =0$, it must be nonnegative, which reduces
to the previous case.
\hfill \framebox{}

%It is easy to verify that Corollary \ref{prop.pinskerf.easy}
%implies that $\chi^2 \geq V^2$ and $\Delta \geq \frac{1}{2} V^2$,
%which we have already mentioned.
%For the Hellinger discrimination we have that
%$\tilde{f} (u) = \frac{1}{2} (\sqrt{u}-1)^2$, $f''(1)=\frac{1}{4}$ and
%$w_f = \frac{1}{2}$.
%Since the left hand side of (\ref{cond.2nd.easy}) is $\frac{3}{4} \frac{|u-1|}{u}$,
%we conclude that $h^2 \geq \frac{1}{8} V^2$
%(this is weaker than
%$4 h^2 (2 - h^2) \geq V^2$, which we already mentioned and
%implies that $h^2 \geq 1 - \sqrt{1 - V^2 /4} \geq V^2 /8$).
In our view  the most important application of
Corollary \ref{prop.pinskerf.easy} is to the
relative information of type $(1 - \alpha)$ and the
R\'{e}nyi's information gain of order $\alpha$.

\begin{corollary}
$D_{(\alpha)} \geq \frac{1}{2} \, V^2$
whenever $-1 \leq \alpha \leq 2$, $\alpha \not= 0,1$.
Also, ${\cal I}_\alpha \geq \frac{1}{2} \, \alpha \, V^2$
for $0< \alpha < 1$.
In both cases the coefficient of $V^2$ is best possible.
\end{corollary}

{\sc Proof:}
For $D_{(\alpha)}$ we have
$f(u) = [\alpha (\alpha -1)]^{-1}(u^{\alpha} -1)$.
It is easy to check that $f''(1) = 1$, $w_f = (\alpha+1)/3$ and
the left hand side of (\ref{cond.2nd.easy}) is
$(\alpha +1) (2 - \alpha) |u-1| / 3u$, which satisfies
the condition for $-1 \leq \alpha \leq 2$.

Now, for $0 < \alpha < 1$, write
${\cal I}_{\alpha} = (1-\alpha)^{-1} \log \left[ 1 + \frac{\alpha (1 - \alpha) D_{(1-\alpha)}}{1 - \alpha (1 - \alpha) D_{(1-\alpha)}} \right]$
and use that $\log (1+x) \geq \frac{2 x}{2+x}$ for $x \geq 0$
(cf.\ Tops{\o}e \cite{topsoelog04})
to get that
${\cal I}_{\alpha} \geq \alpha D_{(1-\alpha)} \, \frac{2}{2 - \alpha (1 - \alpha) D_{(1-\alpha)}} \geq \frac{1}{2} \, \alpha \, D_{(1-\alpha)}$.
\hfill \framebox{}

{\sc Remark 1}. Pinsker's inequality can be seen as the
limiting case of the inequality just stated for
$D_{(\alpha)}$ as $\alpha \to 0$ or equivalently
of that stated for ${\cal I}_{\alpha}$ as $\alpha \uparrow 1$.

{\sc Remark 2}. The behavior of ${\cal I}_{\alpha}$ for $\alpha > 1$
is somewhat puzzling to us.
We will show in Section 4
(cf.\ Theorem \ref{prop.pinsker.f4} and Corollary \ref{prop.relinfo4})
that for any $-1 \leq \alpha \leq 2$
there are probability measures $P_v$ and $Q_v$ such that
$V(P_v,Q_v) = v$ and
$D_{(1-\alpha)} (P_v,Q_v) = \frac{1}{2} \, v^2 +
\frac{1}{72} \, (\alpha+1)(2-\alpha) \, v^4 + o(v^4)$.
Hence, we must have that
${\cal I}_{\alpha} (P_v , Q_v) =
(\alpha -1)^{-1} \log [1-\alpha (1-\alpha) D_{(1-\alpha)} (P_v,Q_v)] =
\frac{1}{2} \alpha v^2 +
\frac{1}{36} \alpha (1+5 \alpha - 5 \alpha^2) v^4 + o (v^4)$.
Since $1+5 \alpha - 5 \alpha^2 < 0$ for
$\alpha > \frac{1}{2} + \frac{3}{10} \, \sqrt{5} \approx 1.17$,
it follows that in this case we cannot have that
${\cal I}_\alpha \geq \frac{1}{2} \alpha V^2$.
We do not know whether the inequality holds for
$1 < \alpha \leq \frac{1}{2} + \frac{3}{10} \, \sqrt{5}$.

\section{Fourth-order inequalities}

The inequality $D \geq \frac{1}{2} V^2 + \frac{1}{36} V^4$ is
a consequence of the fact that
\begin{equation}
\label{ineq.log.4}
u-1- \log u \geq \frac{1}{2} \, \frac{(u-1)^2}{1 + \frac{2}{3} (u-1)} +
\frac{1}{36} \, \frac{(u-1)^4}{[1 + \frac{28}{45} (u-1)]^3}
\end{equation}
for all $u > 0$ together with (\ref{eqn.lemma.2}) and (\ref{eqn.lemma.4}).
To prove (\ref{ineq.log.4}), let
$g(u) = (u-1- \log u ) [1 + \frac{2}{3} (u-1)] [1 + \frac{28}{45} (u-1)]^3 -
\frac{1}{2} (u-1)^2 [1 + \frac{28}{45} (u-1)]^3 - \frac{1}{36} (u-1)^4 [1 + \frac{2}{3} (u-1)]$
and use Lemma \ref{lemma.derivs} after showing that
$g(1) = g'(1) = g''(1) = g'''(1) = g^{(4)} (1) = g^{(5)} (1) = 0$ and
$g^{(6)} (u) = \frac{8}{91125}
(43904 u^4-50960 u^3 +44268 u^2-34102 u + 24565)/u^6$
is positive everywhere because
\begin{eqnarray}
\lefteqn{43904 u^4-50960 u^3 +44268 u^2-34102 u + 24565} \nonumber \\
& & =
43904 (u-\frac{65}{224})^4 +
\frac{88347}{4} (u-\frac{1907903}{2827104})^2  +
\frac{10273158845617}{723738624} \,.
\label{eqn.deriv6.log}
\end{eqnarray}

In this section we generalize the idea in the last paragraph for arbitrary
$f$-divergences. In other words, an inequality of the form
$D_f \geq c_{2,f} V^2 + c_{4,f} V^4$ would be obtained
if we can prove that
\begin{equation}
\label{eqn.f4}
\tilde{f} (u) \geq c_{2,f} \frac{(u-1)^2}{1+(1-w_{2,f})(u-1)} +
c_{4,f} \frac{(u-1)^4}{[1+(1-w_{4,f})(u-1)]^3}.
\end{equation}
for all $u>0$.
For this inequality to hold and being sharp enough
so that $c_{2,f}$ and $c_{4,f}$ are best possible,
it is necessary
that the Taylor expansions of both sides around $u=1$
must coincide up to and including fifth-order terms.
This condition implies the expression of the
$c_{i,f}$'s and $w_{i,f}$'s ($i = 2,4$) in terms
of the derivatives of $f$ at $u=1$
in the next theorem.

\begin{theorem}
\label{prop.pinsker.f4}
{\bf (Fourth-order extended Pinsker inequality for $f$-divergences)}.
Let $f$ and $\tilde{f}$ be as before and define $c_{2,f} = \frac{f''(1)}{2}$,
$w_{2,f} = 1+ \frac{1}{3} \frac{f'''(1)}{f''(1)}$,
$c_{4,f} = \frac{1}{72} [3 f^{(4)} (1) - 4 \frac{(f''')^2 (1)}{f''(1)} ]$
and 
\[
w_{4,f} = 1 + \frac{1}{45} \,
\frac{9 f^{(5)} (1) - 20 \frac{(f''')^3 (1)}{(f'')^2 (1)}}{
3 f^{(4)} (1) - 4 \frac{(f''')^2 (1)}{f''(1)}} \,.
\]
Suppose that both $c_{2,f} > 0$ and $c_{4,f} > 0$ 
and that for every $u > 0$ we have
\begin{eqnarray}
\label{cond.4th}
\lefteqn{
\tilde{f} (u) \, [1 + (1-w_{2,f})(u-1)] \, [1 + (1-w_{4,f})(u-1)]^3} \nonumber \\
& & \geq
c_{2,f} \, (u-1)^2  \, [1 + (1-w_{4,f})(u-1)]^3 +
c_{4,f} \, (u-1)^4  \, [1 + (1-w_{2,f})(u-1)] \,.
\end{eqnarray}
Then $D_f \geq c_{2,f} V^2 + c_{4,f} V^4$.
The coefficient $c_{4,f}$ is best possible.
\end{theorem}

{\sc Proof:}
We will prove first that (\ref{cond.4th}) implies that
$0 \leq w_{i,f} \leq 1$ ($i=2,4$). Reasoning by contradiction,
suppose first that $w_{2,f} \not\in [0,1]$, and evaluate
both sides of (\ref{cond.4th}) at $u = -w_{2,f}/(1-w_{2,f}) > 0$ to obtain
that $0 \geq c_{2,f} (1-w_{2,f})^5 (w_{4,f}-w_{2,f})^3$.
Since $c_{2,f} > 0$, this implies that
$w_{4,f} \geq w_{2,f}$ when $w_{2,f} > 1$ and
$w_{4,f} \leq w_{2,f}$ when $w_{2,f} < 0$.
Hence also $w_{4,f} \not\in [0,1]$ and we can
evaluate again both sides of (\ref{cond.4th}) now at
$u = -w_{4,f}/(1-w_{4,f}) > 0$ to get that
$0 \geq c_{4,f} (1-w_{4,f})^5 (w_{2,f}-w_{4,f})$,
so that $w_{4,f} > 1$ implies now that $w_{2,f} \geq w_{4,f}$
while $w_{4,f} < 0$ implies that $w_{2,f} \leq w_{4,f}$.
Hence we must have in any case that
$w_{2,f}$ and $w_{4,f}$ are equal, say
$w_{2,f} = w_{4,f} =w$, so that (\ref{cond.4th}) becomes
\begin{equation}
\label{ineq.temp}
\tilde{f} (u) \, [1 + (1-w)(u-1)]^4 \geq
c_{2,f} \, (u-1)^2  \, [1 + (1-w)(u-1)]^3 +
c_{4,f} \, (u-1)^4  \, [1 + (1-w)(u-1)] \,.
\end{equation}
Since we are still assuming that $w_{2,f} = w \not\in [0,1]$,
consider this inequality as
$u \to -w/(1-w) > 0$.
The left hand side is equivalent to
$f [-w/(1-w)] (1-w)^4 [u + w/(1-w)]^4$, {\em i.e.}\
a positive coefficient times an infinitesimal
term of order 4 in $[u + w/(1-w)]$,
while the right hand side is equivalent to
$c_{4,f} (1-w)^3 [u + w/(1-w)]$,
{\em i.e.}\ an infinitesimal
term of order 1 in $[u + w/(1-w)]$.
If $w > 1$ we have that the principal part
here is $c_{4,f} (1-w)^3 < 0$ and hence (\ref{ineq.temp})
cannot hold as $u \uparrow -w/(1-w)$,
while if $w <0$ the principal part of the right hand side
is $c_{4,f} (1-w)^3 > 0$ and (\ref{ineq.temp})
cannot hold as $u \downarrow -w/(1-w)$.
This contradiction is due to the assumption
that $w_{2,f} \not\in [0,1]$.
Similarly we prove that $0 \leq w_{4,f} \leq 1$.

Now that we have proved that both
$[1 + (1-w_{2,f})(u-1)]$ and $[1 + (1-w_{4,f})(u-1)]$
are positive for $u > 0$, we observe that
Condition (\ref{cond.4th}) implies (\ref{eqn.f4})
and hence that
\begin{eqnarray*}
\lefteqn{D_f (P,Q) = D_{\tilde{f}} (P,Q)} \\
& & \geq
c_{2,f} \int \frac{(q/p-1)^2}{1+(1-w_{2,f})(q/p-1)} \, p \, d \mu +
c_{4,f} \int \frac{(q/p-1)^4}{[1+(1-w_{4,f})(q/p-1)]^3} \, p \, d \mu \,.
\end{eqnarray*}
Hence, use respectively
(\ref{eqn.lemma.2}) and (\ref{eqn.lemma.4})
to bound each term in the right hand side to obtain
that $D_f \geq c_{2,f} \, V^2 + c_{4,f} \, V^4$.
%\[
%\frac{(u-1)^4}{[\tilde{f} (u) - c_{2,f} (u-1)^2 / (1+(1-w_{2,f}) (u-1))]
%[1 + (1-w_{4,f}) (u-1)]^3} \leq c_{4,f}^{-1} \,.
%\]
%$f^*_0 (u) = \tilde{f} (u) - c_{2,f} (u-1)^2 / (1+(1-w_{2,f}) (u-1))$
%({\bf Abuso de notacao!!!!!!!})
%and $k(u) = 1 + (1-w_{4,f}) (u-1)$ to obtain that
%\[
%D_{f^*} = D_f - c_{2,f} \int \frac{(q/p-1)^2}{1+(1-w_{2,f}) (q/p-1)} p \geq
%c_{4,f} V^4 \, .
%\]
%Finally, take
%$f^*_0 (u) = (u-1)^2 / [1+(1-w_{2,f}) (u-1)]$ and
%$k(u) = 1+(1-w_{2,f}) (u-1)$
%in Lemma \ref{eqn.lemma.2} to obtain that
%$\int \frac{(q/p-1)^2}{1+(1-w_{2,f}) (q/p-1)} p \geq V^2$.

The proof that $c_{4,f}$ is best possible is
similar to the last part of the proof of
Theorem \ref{prop.pinskerf}.
Consider again a binary space and for small $v$ define
$P = (p, 1-p)$ and $Q_v = (p+v/2, 1-p-v/2)$ so that
$V(P,Q_v) = v$ and $D_f (P,Q_v) = p f(1+v/2p) + (1-p) f (1-v/2(1-p))$.
Now let $p= \frac{1}{2} + \frac{1}{6} \frac{f''' (1)}{f'' (1)} v$ and
expand $D_f (P,Q_v)$ around $v=0$ to obtain that
$D_f (P,Q) = c_{2,f} \, v^2 + c_{4,f} \, v^4 + o (v^4)$.
We leave the details to the reader.
\hfill \framebox{}

The following condition on the derivatives of $f$ is usually easier
to prove than (\ref{cond.4th}).

\begin{corollary}
\label{prop.pinskerf4.easy}
Let $f$, $\tilde{f}$, $c_{2,f}$, $w_{2,f}$, $c_{4,f}$ and $w_{4,f}$ be as before,
and suppose that $f$ is six times differentiable with
$f''(u) > 0$ for all $u$. Then
\begin{eqnarray}
\label{cond.4th.easy}
\lefteqn{
\frac{f^{(6)} (u)}{f'' (u)} [1+(1-w_{2,f})(u-1)] [1+(1-w_{4,f})(u-1)]^3
} \nonumber \\
& & +
6 \frac{f^{(5)} (u)}{f'' (u)} [1+(1-w_{4,f})(u-1)]^2 [4 - w_{2,f} - 3 w_{4,f}
+ 4 (1-w_{2,f})(1-w_{4,f}) (u-1) ]
\nonumber \\
& & +
90 \frac{f^{(4)} (u)}{f'' (u)} (1- w_{4,f}) [1+(1-w_{4,f})(u-1)]
[2 - w_{2,f} - w_{4,f} + 2 (1-w_{2,f})(1-w_{4,f}) (u-1) ]
\nonumber \\
& & +
120 \frac{f''' (u)}{f'' (u)} (1- w_{4,f})^2
[4 - 3 w_{2,f} - w_{4,f} + 4 (1-w_{2,f})(1-w_{4,f}) (u-1) ]
\nonumber \\
& & +
144 (1- w_{2,f}) (1- w_{4,f})^3
\geq 0
\end{eqnarray}
implies (\ref{cond.4th}) and hence that
$D_f \geq c_{2,f} \, V^2 + c_{4,f} \, V^4$.
\end{corollary}

{\sc Proof:}
Let $g(u) = \tilde{f} (u) [1+(1-w_{2,f})(u-1)] [1+(1-w_{4,f})(u-1)]^3
- c_{2,f} (u-1)^2 [1+(1-w_{4,f})(u-1)]^3 - c_{4,f} (u-1)^4 [1+(1-w_{2,f})(u-1)]$.
Now it is straightforward, although rather tedious,
that $g(1) = g'(1) = g''(1) = g'''(1) = g^{(4)} (1) = g^{(5)} (1) =0$,
while $g^{(6)} (u)$ equals the left hand side of (\ref{cond.4th.easy}) times
$f'' (u)$ and hence is nonnegative.
%(Note also from the definition of $f$ and $\tilde{f}$ that
%the second and higher-order derivatives of $f$ and $\tilde{f}$ coincide).
Hence, Lemma \ref{lemma.derivs} implies that $g(u) \geq 0$, which
is equivalent to (\ref{cond.4th}).
\hfill \framebox{}

The case of Jeffrey's divergence
($f(u) = \tilde{f} (u) = (u-1) \log u$) is interesting,
because we have that
$c_{2,f} = 1$, $w_{2,f} = \frac{1}{2}$,
$c_{4,f} = \frac{1}{12}$, $w_{4,f} = \frac{1}{2}$ and
the left hand side of (\ref{cond.4th.easy}) is
$\frac{3}{2}
(5 u^4 -8 u^3 +9 u^2 -8 u +5)/u^4 = \frac{3}{2}
[5 (u- \frac{2}{5})^4 + \frac{21}{5} (u - \frac{4}{5})^2 + \frac{273}{125}]/u^4 >0$.
Hence $J \geq V^2 + \frac{1}{12} V^4$.
We note that from $D \geq \frac{1}{2} V^2 + \frac{1}{36} V^4$
it follows immediately using the symmetry of $V$ that
$J (P,Q) = D(P,Q) + D(Q,P) \geq V^2 + \frac{1}{18} V^4$,
but this bound is worst than the one we found using
Corollary \ref{prop.pinskerf4.easy}.
To finish this section we present the special case of
the relative information and information gain of order $\alpha$.

\begin{corollary}
\label{prop.relinfo4}
$D_{(\alpha)} \geq \frac{1}{2} V^2 +
\frac{1}{72} (\alpha+1)(2-\alpha) V^4$
for $-1 \leq \alpha \leq 2$, $\alpha \not= 0,1$.
Also, ${\cal I}_{\alpha} \geq \frac{\alpha}{2} V^2 + \frac{1}{36} \alpha (1 + 5 \alpha - 5 \alpha^2 ) V^4$
for  $0 < \alpha < 1$.
In both cases the coefficients of $V^4$ are best possible.
\end{corollary}

{\sc Proof:}
We prove first the assertion for $D_{(\alpha)}$.
We have $f(u) = [\alpha (\alpha-1)]^{-1} (u^{\alpha} -1)$
and then $c_{2,f} = \frac{1}{2}$, $w_{2,f} = \frac{\alpha+1}{3}$,
$c_{4,f} = \frac{1}{72} (\alpha+1)(2-\alpha)$,
$w_{4,f} = \frac{17 + 11 \alpha}{45}$. Again
straightforwardly although rather lengthy,
the left hand side of (\ref{cond.4th.easy}) is
\begin{eqnarray}
\label{ineq.alpha.fourth}
\lefteqn{
\frac{(\alpha+1)(2-\alpha)}{273375 \, u^{-4}}
\left[
- (\alpha-5) (\alpha-3) (\alpha-4) (11\alpha+17)^3 \right.}
\nonumber \\ & & +
2 (\alpha-3) (\alpha-4) (22\alpha^2-28\alpha-59) (11 \alpha + 17)^2 u
\nonumber \\ & & -
6 (11 \alpha + 17) (-28+11 \alpha) (\alpha-3)(\alpha+2)(11 \alpha^2 - 11 \alpha - 31) u^2
\nonumber \\ & & +
2 (3+\alpha) (\alpha+2) (22 \alpha^2 - 16 \alpha - 65) (-28+11 \alpha)^2 u^3
\nonumber \\ & & - \left.
(\alpha+4) (3+\alpha) (\alpha+2) (-28+11 \alpha)^3 u^4
\right] \,,
\end{eqnarray}
which we prove in the appendix that is positive for $u > 0$
when $-1 \leq \alpha \leq 2$.

Now, for R\'{e}nyi's information gain we proceed as in the proof of Corollary \ref{prop.pinskerf.easy}
and use again that $\log (1+x) \geq \frac{2 x}{2+x}$ for $x \geq 0$, so that
\begin{eqnarray*}
\lefteqn{
{\cal I}_{\alpha} = (1-\alpha)^{-1}
\log \left[ 1 + \frac{\alpha (1 - \alpha) D_{(1-\alpha)}}{1 - \alpha (1 - \alpha) D_{(1-\alpha)}} \right] } \\
& & \geq
\alpha D_{(1-\alpha)} \, \frac{2}{2 - \alpha (1 - \alpha) D_{(1-\alpha)}} \geq
\alpha D_{(1-\alpha)} [1 + \frac{1}{2} \alpha (1-\alpha) D_{(1-\alpha)}] \\
& & \geq
\frac{1}{2} V^2 +
\frac{1}{72} (\alpha+1)(2-\alpha) V^4 + \frac{\alpha^2}{2}(1-\alpha) \frac{1}{4} V^4 =
\frac{\alpha}{2} V^2 + \frac{1}{36} \alpha (1 + 5 \alpha - 5 \alpha^2 ) V^4 \,.
\end{eqnarray*}
\hfill \framebox{}

Using Corollary \ref{prop.relinfo4}
we can obtain bounds for the 
Tsallis and Cressie-Read divergences, 
since they can be expressed in terms of $D_{(\alpha)}$. 

%{\bf Hellinger discrimination}.
%We have $\tilde{f} (u) = \frac{1}{2} (\sqrt{u}-1)^2$. Hence
%$w_f = \frac{1}{4}$ and
%the left hand side of (\ref{cond.2nd.easy}) is $\frac{3}{4} \frac{|u-1|}{u}$.
%Hence $h^2 \geq \frac{1}{8} V^2$, which in any case is weaker than
%$4 h^2 (2 - h^2) \geq V^2$, which we already mentioned and
%implies that $h^2 \geq 1 - \sqrt{1 - V^2 /4} \geq V^2 /8$.
%\item
%{\bf $\chi^2$ divergence}.
%$\tilde{f} (u) = (u-1)^2$, $w_f = 1$ and
%the left hand side of (\ref{cond.2nd.easy}) is $0$, hence $\chi^2 \geq V^2$.
%\item
%{\bf Triangular discrimination}.
%$\tilde{f} (u) = (1-u)^2 / (1+u)$, $w_f = 1/2$ and
%the left hand side of (\ref{cond.2nd.easy}) is $0$, hence $\Delta \geq \frac{1}{2} V^2$,
%which we have already mentioned.

\section{A note about higher-order inequalities}

Unfortunately, the tools we have used so far are
insufficient to obtain inequalities of the form
$D_f \geq c_{2,f} V^2 + c_{4,f} V^4 +
\cdots + c_{2n,f} V^{2n}$ including terms of at least order $6$.
To understand why, it is useful to
restrict attention to the information divergence $D$.
Our discussion at the beginning of Sections 2 and 3
shows that the second and fourth-order inequalities
for $D$ are due respectively to the inequalities
$u-1- \log u \geq \frac{1}{2} (u-1)^2 / [1 + \frac{2}{3} (u-1)]$ and
(\ref{ineq.log.4}) and Corollary \ref{prop.lemma.vmn}.
Now, it is straightforward to check that the difference
between the left and the right hand sides of (\ref{ineq.log.4})
equals $\frac{41}{12150} (u-1)^6 + O(|u-1|^7)$.
Indeed, we conjecture that
\begin{equation}
\label{ineq.log.6}
u-1 - \log u \geq \frac{1}{2} \, \frac{(u-1)^2}{1 + \frac{2}{3} (u-1)}
+ \frac{1}{36} \, \frac{(u-1)^4}{[1 + \frac{28}{45} (u-1)]^3}
+ \frac{41}{12150} \, \frac{(u-1)^6}{[1+\frac{23186}{38745} (u-1)]^5} \, .
\end{equation}
However, even if we could prove this assertion,
we would obtain then only that
$D \geq \frac{1}{2} V^2 + \frac{1}{36} V^4 + \frac{41}{12150} V^6$.
The coefficient $\frac{41}{12150}$, even if close, is smaller than
the best possible $\frac{1}{270}$ found by
Tops{\o}e \cite{topsoe2e100}.
A possible explanation
for this, or in other words, from where
the difference $\frac{1}{270} - \frac{41}{12150} = \frac{2}{6075}$
comes from, can be obtained looking at the
divergences which result from the
right hand side of (\ref{ineq.log.6}).
We have on one side that
$\int \frac{(q/p-1)^2}{1+(2/3)(q/p-1)} \, p \, d \mu \geq V^2(P,Q)$,
with equality holding if and only if
$[1 + \frac{2}{3} (q/p-1)] \propto |q/p-1|$
(actually, the proportionality constant must equal $1/V(P,Q)$).
Similarly,
$\int \frac{(q/p-1)^4}{[1+(28/45)(q/p-1)]^3} \, p \, d \mu \geq V^4(P,Q)$,
with equality holding if and only if
$[1 + \frac{28}{45} (q/p-1)] \propto |q/p-1|$.
Hence, it follows from these two inequalities that
\begin{equation}
\label{eqn.surplus}
\frac{1}{2}
\int \frac{(q/p-1)^2}{1+(2/3)(q/p-1)} \, p \, d \mu +
\frac{1}{36}
\int \frac{(q/p-1)^4}{[1+(28/45)(q/p-1)]^3} \, p \, d \mu \geq
\frac{1}{2} V^2(P,Q) + \frac{1}{36} V^4(P,Q) \,,
\end{equation}
but now equality cannot hold since, even if close,
$\frac{2}{3} \not= \frac{28}{45}$.
In fact, we conjecture that the infimum of the left
hand side of (\ref{eqn.surplus}) taken over all
$P$ and $Q$ such that $V(P,Q) = v$ equals
$\frac{1}{2} \, v^2 + \frac{1}{36} \, v^4 +
(\frac{1}{270} - \frac{41}{12150}) v^6 + o(v^6)$.
We hope to be able to report on these issues soon.

\section{Appendix}

Before proving (\ref{ineq.alpha.fourth}) we will
state the following lemma.
We have already used the idea in the lemma
to obtain the decomposition (\ref{eqn.deriv6.log}).

\begin{lemma}
\label{lemma.poli4}
Let $T(u) = c_4 u^4 + c_3 u^3 + c_2 u^2 + c_1 u + c_0$
and define $a_4 = c_4$, $a_2 = \frac{1}{8} \, [8 c_2 c_4 - 3 c_3^2]/c_4$ and
\begin{equation}
\label{eqn.lemma.a0}
a_0 = \frac{1}{256 c_4^3} \, \frac{
2048 c_0 c_4^4 c_2 - 768 c_0 c_4^3 c_3^2 - 8 c_3^4 c_2 c_4 +
c_3^6 + 64 c_3^3 c_1 c_4^2 - 512 c_1^2 c_4^4}{
8 c_2 c_4 - 3 c_3^2} \,.
\end{equation}
Then a sufficient condition for $T(u) \geq 0$ for every $u$ is that
$a_4$, $a_2$ and $a_0$ are nonnegative.
\end{lemma}

{\sc Proof:}
Check that
\[
T(u) = a_4 \, (u+\frac{c_3}{4 c_4} )^4 +
a_2 \, (u + \frac{1}{4 c_4} \, \frac{16 c_1 c_4^2 - c_3^3}{8 c_2 c_4 - 3 c_3^2})^2
+ a_0 \,.
\]
\hfill \framebox{}

{\sc Proof of (\ref{ineq.alpha.fourth})}:
Let $-1 \leq \alpha \leq 2$,
$T(u) = T_{\alpha} (u)$ be the term between brackets in (\ref{ineq.alpha.fourth}) and
define $c_i = c_i (\alpha)$ and $a_i = a_i (\alpha)$ as in
Lemma \ref{lemma.poli4}, so that for instance
$a_4 = c_4 = - (\alpha+4) (\alpha+3) (\alpha+2) (11 \alpha -28)^3$
which is of course nonnegative. Next
\[
a_2 = \frac{9}{2} \frac{(980 \alpha^3 + 552 \alpha^2 - 4257 \alpha - 4207)
(\alpha + 2) (-28 + 11 \alpha)}{\alpha+4} \,,
\]
which is positive because
\[
980 \alpha^3 + 552 \alpha^2 - 4257 \alpha - 4207 =
-(980 \alpha + 1532) (2-\alpha) (\alpha+1)- (1143+765 \alpha)
\]
is negative. 
%Hence to conclude the proof we
%need to show that $a_0 \geq 0$.
Using (\ref{eqn.lemma.a0}) we obtain that
$a_0 = 9 P_{10} (\alpha) / 32 a_2 (\alpha +4)^4$ where
\begin{eqnarray*}
\lefteqn{P_{10} (\alpha) = -
20792743232 \alpha^{10} - 168248775872 \alpha^9 + 54551858544 \alpha^8} \\
& & +
3066837388032 \alpha^7 + 4844633801556 \alpha^6 -
14799467270700 \alpha^5 - 43681339670379 \alpha^4 \\
& & -
4381425810042 \alpha^3 + 94728169651149 \alpha^2 +
113143847999692 \alpha + 41092635382468 \,.
\end{eqnarray*}
Hence, to conclude the proof we need to show that 
$a_0 \geq 0$ or equivalently that 
$P_{10} (\alpha) \geq 0$
whenever $-1 \leq \alpha \leq 2$.
This follows from the following identity
\begin{eqnarray}
\lefteqn{P_{10} (\alpha) =
(20792743232 \alpha^2+189041519104 \alpha+300831606416)(2-\alpha)^3 (\alpha+1)^5}
\nonumber \\
& & + (1295259115248 \alpha + 3335882569236) (2 - \alpha)^2 (\alpha+1)^4 \nonumber \\
& & +
(1471491213228 \alpha + 7953881034231) (2-\alpha) (\alpha+1)^3 \nonumber \\
& & + 1343948812407 (2-\alpha) (\alpha+1)^2 \nonumber \\
& &  + (4661891728632 \alpha^2 + 11252369540556 \alpha + 6746792560920) \,,
\label{eqn.p10}
\end{eqnarray}
since after examining the signs of the different factors it is possible
to conclude that each term in the sum is nonnegative.
\hfill \framebox{}

{\sc Remark}. Checking that this last identity holds constitutes
a formal proof of the fact that $P_{10} (\alpha) \geq 0$ for
every $-1 \leq \alpha \leq 2$. Explaining how we obtain it
is a bit harder, specially since it involves a "trial and error"
process. Essentially, we try to divide $P_{10} (\alpha)$
by a polynomial $A(\alpha)$ of degree $a$ which was known to be positive for
the desired range. Hence we obtain that
$P_{10} (\alpha) = Q(\alpha) A(\alpha) + R(\alpha)$, where
the degrees of $Q$ and $R$ are at most $(10-a)$ and $(a-1)$.
A sufficient condition for $P_{10} (\alpha) \geq 0$ is then that
both $Q(\alpha)$ and $R(\alpha)$ are nonnegative for the desired
range.
Since we have $-1 \leq \alpha \leq 2$,
natural candidates for $A ( \alpha)$ took the form
$(2-\alpha)^m (\alpha+1)^n$.
The polynomial division can be made easily using a
symbolic manipulation package (cf. the function {\tt quo} in MAPLE),
while a plotting routine can make an initial assessment
of whether the decomposition was successful, i.e.\ whether
both $Q$ and $R$ are nonnegative
(if not, we would try again with different $m$ and $n$).
For instance, the first successful division made to arrive to
(\ref{eqn.p10}) had $A ( \alpha ) = (2-\alpha)^3 (1+\alpha)^5$.
In a sense this means that we change a degree $10$ problem
(showing that $P_{10}$ is nonnegative) by two problems
having degrees 2 and 7 (showing respectively that
$Q$ and $R$ are nonnegative).
The same procedure can be repeated for $Q$ and for $R$ and
then for their respective quotients and rests and so on until
all polynomials involved are either of the form
$(2-\alpha)^m (\alpha+1)^n$ or have at most degree 2 and
their roots (hence their signs) can be obtained analytically.
After doing all these divisions it is easy to put back everything
together into a unique decomposition as in (\ref{eqn.p10}).
Of course, we have no guarantee that this procedure would
work for any polynomial, but it worked for $P_{10}$.

\mbox{}

\noindent
{\bf Acknowledgements}

\mbox{}

This research was partially funded by
a CAPES-PROCAD grant.
The author is also grateful to the Department
of Mathematics of the Universit\`{a} di Roma "La Sapienza"
and specially to Prof. F.\ Spizzichino for
support during a sabbatical leave.

\bibliography{}%../information}

\begin{thebibliography}{10}

\bibitem{VoRo59}
V.~A. Volkonskij and J.~A. Rozanov, ``Some limit theorems for random functions
  - {I},'' {\em Theory Prob. Appl.}, vol.~4, pp.~178 -- 197, 1959.

\bibitem{fedotovetal03}
A.~Fedotov, P.~Harremo{\"e}s, and F.~Tops{\o}e, ``Refinements of {P}insker's
  {I}nequality,'' {\em IEEE Trans. Inf. Theory}, vol.~49, pp.~1491--1498, June
  2003.

\bibitem{barron86}
A.~R. Barron, ``Entropy and the {C}entral {L}imit {T}heorem,'' {\em Annals
  Probab.}, vol.~14, no.~1, pp.~336 -- 342, 1986.

\bibitem{topsoe79}
F.~Tops{\o}e, ``Information theoretical optimization techniques,'' {\em
  Kybernetika}, vol.~15, no.~1, pp.~8 -- 27, 1979.

\bibitem{harruzankin04}
P.~Harremo{\"e}s and P.~Ruzankin, ``Rate of convergence to {P}oisson law in
  terms of information divergence,'' {\em IEEE Trans. Inf. Theory}, vol.~50,
  no.~9, pp.~2145--2149, 2004.

\bibitem{delpino02}
M.~{Del Pino} and J.~Dolbeault, ``Best constants for {G}agliardo-{N}iremberg
  inequalities and applications to nonlinear difussions,'' {\em Journal
  Mathematiques Pures et Appliquees}, vol.~81, pp.~847--875, 2002.

\bibitem{Csiszar67}
I.~Csisz{\'a}r, ``Information-type measures of difference of probability
  distributions and indirect observations,'' {\em Studia Sci. Math. Hungar.},
  vol.~2, pp.~299--318, 1967.

\bibitem{Kemperman69}
J.~H.~B. Kemperman, ``On the optimal rate of transmitting information,'' {\em
  Ann. Math. Statist}, vol.~40, pp.~2156 -- 2177, Dec. 1969.

\bibitem{Pinsker64}
M.~S. Pinsker, {\em Information and Information Stability of Random Variables
  and Processes}.
\newblock A. Feinstein, tr. and ed., San Francisco: Holden-Day, 1964.

\bibitem{Kean66}
H.~P. {McKean Jr.}, ``Speed of approach to equilibrium for {K}ac's caricature
  of a {M}axwellian gas,'' {\em Arch. Rational Mech. Anal.}, vol.~21, pp.~343
  -- 367, 1966.

\bibitem{Csiszar66}
I.~Csisz{\'a}r, ``A note on {J}ensen's inequality,'' {\em Studia Sci. Math.
  Hungar.}, vol.~1, pp.~185--188, 1966.

\bibitem{Kullback67}
S.~Kullback, ``A lower bound for discrimination information in terms of
  variation,'' {\em IEEE Trans. Inf. Theory}, vol.~IT-13, pp.~126 -- 127, Jan.
  1967.

\bibitem{Kullback70}
S.~Kullback, ``Correction to "a lower bound for discrimination information in
  terms of variation",'' {\em IEEE Trans. Inf. Theory}, vol.~IT-16, p.~652,
  1970.

\bibitem{Vajda70}
I.~Vajda, ``Note on discrimination information and variation,'' {\em IEEE
  Trans. Inf. Theory}, pp.~771--773, Nov. 1970.

\bibitem{topsoe2e100}
F.~Tops{\o}e, ``Bounds for entropy and divergence of distributions over a
  two-element set,'' {\em J. Ineq. Pure Appl. Math.}, vol.~2, Article 25, 2001.

\bibitem{fedotovetal03b}
A.~Fedotov, P.~Harremo{\"e}s, and F.~Tops{\o}e, ``Best {P}insker {B}ound equals
  {T}aylor {P}olynomial of {D}egree $49$,'' {\em Computational {T}echnologies},
  vol.~8, pp.~3--14, 2003.

\bibitem{Ali66}
S.~M. Ali and S.~D. Silvey, ``A general class of coefficients of divergence of
  one distribution from another,'' {\em J. Roy. Statist. Soc. Ser B}, vol.~28,
  pp.~131--142, 1966.

\bibitem{vajda89}
I.~Vajda, {\em Theory of Statistical Inference and Information}.
\newblock London: Kluwer Academic Press, 1989.

\bibitem{taneja04}
I.~Taneja, ``Generalized relative information and information inequalities,''
  {\em J. Ineq. Pure Appl. Math.}, vol.~5(1), Article 21, 2004.

\bibitem{Golan02}
A.~Golan, ``Information and entropy econometrics---editor's view,'' {\em
  Journal of Econometrics}, vol.~107, pp.~1--15, 2002.

\bibitem{Withers99}
L.~{Withers Jr.}, ``Some inequalities relating different measures of divergence
  between two probability distributions,'' {\em IEEE Trans. Inf. Theory},
  vol.~45, no.~5, pp.~1728--1735, 1999.

\bibitem{Csis95}
I.~Csisz{\'a}r, ``Generalized cutoff rates and r{\'e}nyi information
  measures,'' {\em IEEE Trans. Inf. Theory}, vol.~41, pp.~26--34, Jan. 1995.

\bibitem{Dacunha-Castelle78}
D.~Dacunha-Castelle, {\em Ecole d'Ete de Probabilites de Saint-Flour
  VII--1977}.
\newblock Berlin, Heidelberg, New York: Springer, 1978.

\bibitem{Cam86}
L.~{Le Cam}, {\em Asymptotic Methods in Statistical Theory}.
\newblock New York: Springer-Verlag, 1986.

\bibitem{Dragomir01}
S.~S. Dragomir, V.~Glu\u{s}\u{c}evi{\'c}, and C.~E.~M. Pearce, ``{C}sisz{\'a}r
  $f$-divergence, {O}strowski's inequality and mutual information,'' {\em
  Nonlinear Analysis}, vol.~47, pp.~2375--2386, 2001.

\bibitem{Topsoe00ine}
F.~Tops{\o}e, ``Some inequalities for information divergence and related
  measures of discrimination,'' {\em IEEE Trans. Inf. Theory}, vol.~46,
  pp.~1602--1609, 2000.

\bibitem{Csiszar67topological}
I.~Csisz{\'a}r, ``On topological properties of f-divergences,'' {\em Studia
  Sci. Math. Hungar.}, vol.~2, pp.~329--339, 1967.

\bibitem{gilardoni05}
G.~L. Gilardoni, ``Accuracy of posterior approximations via chi-squared and
  harmonic divergences,'' {\em J. of Statistical Planning and Inference},
  vol.~128, no.~2, pp.~475--487, 2005.

\bibitem{gilardoni06}
G.~L. Gilardoni, ``Very accurate posterior approximations based on finite
  mixtures of the hyperparameters conditionals,'' {\em Computational Statistics
  and Data Analysis}, 2006 (to appear).

\bibitem{Kraft55}
C.~Kraft, ``Some conditions for consistency and uniform consistency of
  statistical procedures,'' in {\em Univ. of California Publ. in Statistics,
  vol. 1}, pp.~125--142, Berkeley: Univ. of California, 1955.

\bibitem{Ordentlich05}
E.~Ordentlich and M.~J. Weinberger, ``A distribution dependent refinement of
  {P}insker's inequality,'' {\em IEEE Trans. Inf. Theory}, vol.~51,
  pp.~1836--1840, 2005.

\bibitem{topsoelog04}
F.~Tops{\o}e, ``Some bounds for the logarithmic function,'' in {\em RGMIA
  Research Report Collection}, vol.~7(2), Article 6, 2004.
\newblock To appear in Inequalities Theory and Applications, Vol. 4, Y. J. Cho,
  J. K. Kim and S. S. Dragomir, eds.

\end{thebibliography}
\bibliographystyle{ieeetr}

\end{document}